\documentclass[aps,prb,twocolumn,showpacs,amsmath,amssymb,floatfix]{revtex4-1}
\usepackage{graphicx}
\usepackage{dcolumn}
\usepackage{bm}
\usepackage{amsfonts}
\usepackage{amssymb}
\begin{document}
\title{Monopole and multipole plasmons in a two-dimensional system}
\vspace{2cm}
\author{H. Pfn\"ur}\thanks{e-mail: pfnuer@fkp.uni-hannover.de}
\author{T. Langer}
\author{J. Baringhaus}
\author{C. Tegenkamp}

\affiliation{Institut f\"ur Festk\"orperphysik, Leibniz--Universit\"at Hannover, Appelstrasse 2, D--30167 Hannover, Germany}

\date{\today}

\begin{abstract}
Using monolayer graphene as a model system for a purely two-dimensional (2D) electron gas, we show by energy electron loss spectroscopy, highly resolved both in energy and momentum, that there is a significant probability for the excitation of not only one but two dispersing losses.  The appearance of both losses is independent of the substrate (we tested graphene on the Si face of 6H-SiC(0001), and on Ir(111) without and with intercalated Na layer), and the ratio of the slope in the dispersion curves varies between 1.4 (SiC) and 2.
While the lower dispersion curve can be attributed to the excitation of the monopole plasmon, in
agreement with theoretical model calculations, the upper dispersion branch has not been identified before for plasmonic excitations in a 2D electron gas, and we assign it to the excitation of a multipole sheet plasmon.
\end{abstract}
\pacs{73.22.Lp 73.22.Pr 79.20.Uv} \maketitle

\section{Introduction}

The properties of sheet or wire plasmons  in  low-dimensional  systems \cite{Nagao07}
differ strongly from those in three dimensions. They are
characterized by an acoustic-like dispersion, i.e., their excitation energy
goes to zero in the limit of large wavelengths \cite{Stern67,Allen77,Inaoka02,Ruge08,Ruge10}.  This is due to the lack
of  a  restoring  force for charge density oscillations in the long wavelength limit, contrary
to the behavior of surface plasmons on the surfaces of bulk metals,
which originate simply from the truncation of the bulk electronic system.

These surface plasmons have been identified as self-sustained collective excitations at metal
surfaces \cite{Ritchie57} long ago, but they gained renewed interest in recent years by
investigations of the electromagnetic properties of nanostructured materials \cite{Pendry99,Prodan03} and perspectives
for many potential applications, e.g.
in conjunction with engineering cavity-free and broad band
photon-emitter interactions \cite{Akimov07} via subwavelength confinement
of optical fields near metallic nanostructures \cite{Akimov07,Chang06}. The use of
sheet plasmons in this context, because of their much flatter dispersion compared
with ``conventional'' surface plasmons, would allow an ex-
treme confinement of light in a broad frequency range from medium to far infrared.

In this paper we want to put emphasis on the fact that there is more than one dispersing plasmon mode even
in the plasmons of 2D sheets. This is well known for metallic surfaces of 3D systems, but such a mode
has not been identified for low-dimensional systems yet. For metallic bulk systems, an
additional plasmonic surface mode has first been found theoretically assuming a smoothly decreasing electron density profile
through the surface \cite{Benett70}, which appeared at higher energies than the standard longitudinal mode.
This is the so-called multipole surface plasmon, which was experimentally corroborated only much later
\cite{Schwartz84,Levin79,Tsuei90,Chia00}, mostly on the surfaces of alkali metals and on Al, and by quantitative
DFT calculations \cite{Dobson881,Dobson882,Nazarov01}. It should be pointed out here that the intensity and even
the detectability of the multipole plasmon depends explicitly on the shape of the electronic density distribution normal
to the surface \cite{Liebsch97}.

The adsorption of a 2D conducting sheet, e.g. in form of a graphene monolayer on SiC(0001) or on Ir(111),
represents a localized electron distribution, whose density decays exponentially to both sides normal
to the surface. This decay will be asymmetric because of the screening properties of the substrate, described
by its dielectric function. Therefore, they are also candidates for the existence of multipole plasmons,
but surprisingly, they have not been identified so far. Only asymmetric peak shapes have been detected in most
cases \cite{Liu08,Liu10,Park10}. With simultaneous high energy and momentum resolution, however,
the multipole plasmon loss can be identified, as we will show below.


Due to the low conductance of 1d and 2d metallic systems, one has to cope
with comparatively strong damping of plasmonic excitations
and their sensitivity to defects on the atomic scale. Therefore,
such studies require geometrically perfect and at the same time electronically flexible materials.
Graphene with its two-dimensional lattice of $sp^2$-hybridized carbon atoms has been shown to be an almost ideal
model system for studies of many fundamental aspects of 2d electron systems,
e.g. their electronic correlations, collective phenomena,
many-body interactions, dynamical processes and their
interrelations \cite{Novo04,Geim07,Hebard91,Bost10}. Single layers can
be synthesized by different techniques, e.g. exfoliation of
graphite or decomposition of hydrocarbons on transition metal
surfaces \cite{Novo04,Coraux08, Sutter08}.
SiC substrates by sublimation of Si allow detailed studies of
the morphology, the interface, the electronic structure, and
even of transport properties \cite{Ohta06, Riedl07,Bost08,Haas08,Emtsev08}.
The decay mechanisms of plasmons in graphene layers, the consenquences on the form of the dispersion curve,
and their sensitivity to imperfections  has been subject of several studies by our group \cite{Langer09,Langer10,Tege11}.
While these studies show that the decay mechanisms have to be explicitly taken into account for the understanding
of the dispersion of the plasmons in 2D systems like graphene due to their short plasmonic life times,
we will concentrate here on a detailed line shape analysis of the loss peaks and a comparison of the
results obtained with single graphene sheets on SiC(0001) and on Ir(111) both on the  clean substrate
and intercalated with a submonolayer of Na. This comparison will enable us to see also differences by
the changing position of the Fermi level  in graphene and get new insights into the excitation mechanisms
of low-dimensional plasmons.

\section{Experiment}
We performed all measurements under UHV conditions at a base pressure of $5\times 10^{-9}$Pa.
As substrates, Si-terminated 6H-SiC(0001) samples (n-doped, $\approx 10^{18} $cm$^{-3}$ from
SiCrystal AG) were used, which were etched by an {\em ex-situ} procedure in a furnace of H$_2$
atmosphere in order to remove the residual roughness from polishing steps [29]. Graphene
was grown by sublimation of Si while annealing the SiC substrate to approximately 1500 K and
was stopped after generation of the buffer layer and one additional graphene layer.
Further details can be
found in \cite{Langer09,Langer10}. These procedures resulted large flat terraces with an average
terrace width of about 150 nm, as checked with high resolution LEED \cite{Langer10}.

For graphene on Ir(111)  clean  and well ordered Ir surfaces were generated in a different
vacuum system \cite{Michely08} by repeated cycles of Kr$^+$-sputtering at 1100 K followed by
subsequent flash annealing to 1500 K. Graphene monolayers were grown on these surfaces by
decomposition of hydrocarbons (ethylene) at temperatures around 1320 K, at a pressure of
$2\times10^{-7}$ Pa for 20 sec.
More details are described elsewhere \cite{Michely08}.
These  samples were then transferred into the ELS-LEED systems. Mild annealing to 500 K for several minutes
leads to full desorption of water and other contaminants, as obvious from a
perfect diffraction pattern of a graphene monolayer structure obtained with SPALEED.
The spot profile analysis of the (00) beam in LEED revealed  the terrace width on this surface was approximately
75 nm \cite{Langer11}, which was in good agreement with the terrace structure on bare Ir(111) \cite{Michely08}
For intercalation experiments sodium was deposited from an outgassed dispenser (SAES getters) keeping the pressure in the $10^{-7}$ Pa regime.
\begin{figure}[tb]
 \centering
 \includegraphics[width=0.9\columnwidth]{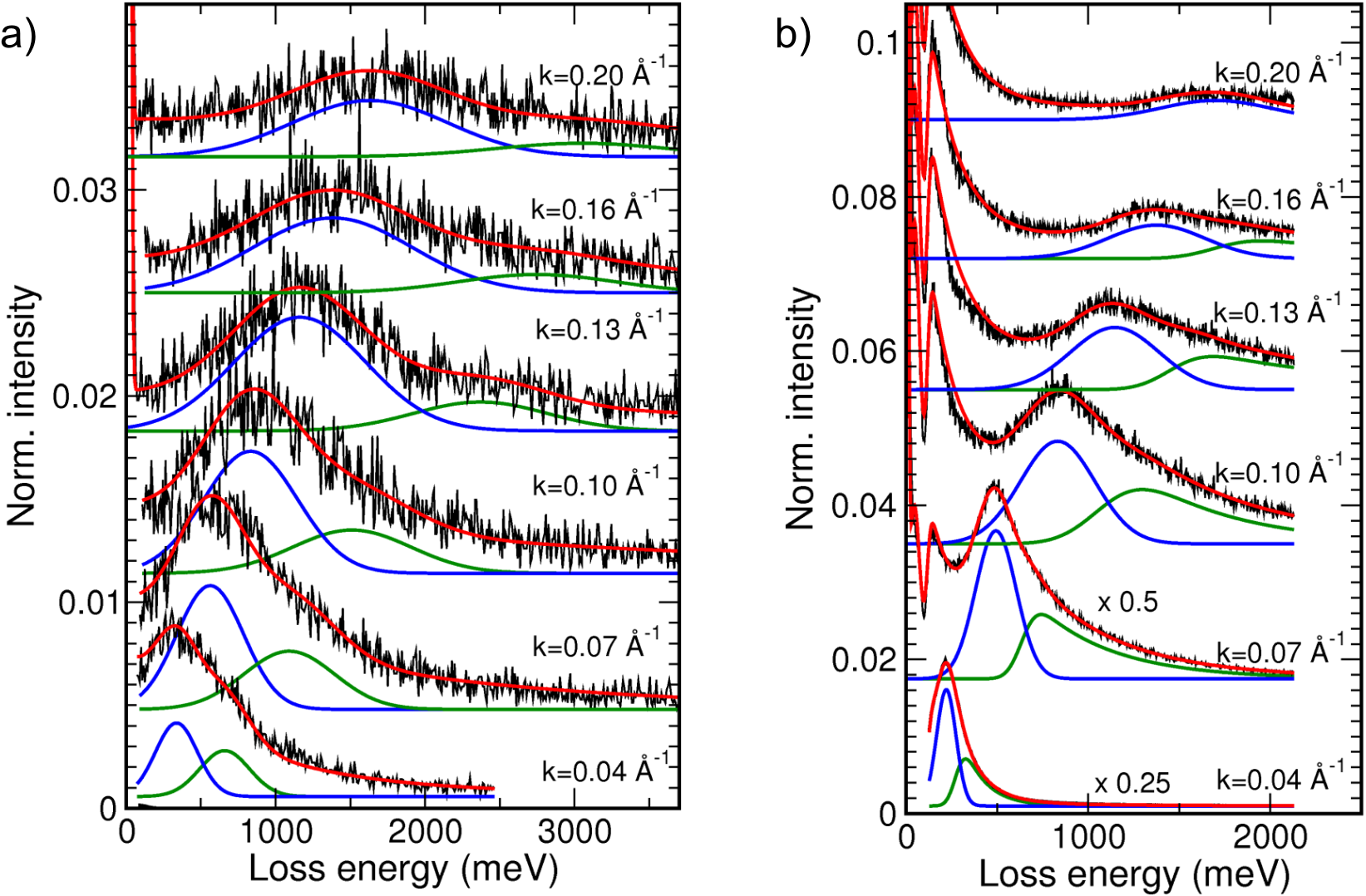}
 \caption{\label{FIG1} Experimental set of EEL spectra as a function of $k_\parallel$.
For monolayer graphene on a) Ir(111), b) SiC(0001)-6H.
Solid curves: Fits of the plasmonic losses of the monopole (blue) and the
multipole plasmons (green). In addition, an exponential form of the Drude tail was fitted (not shown). For graphene on SiC, also the low energy losses due to Fuchs-Kliewer phonons were included in the fits.}
\end{figure}

The plasmons were measured by using
a combination of high resolution electron loss spectrometer
(EELS) as electron source and a detector with a low energy
electron diffraction system (LEED) providing simultaneously
high energy and momentum ($k_\parallel$) resolution [28]. Typical
operating parameters were 25 meV energy resolution at a $k_\parallel$
resolution of $1.3 \times 10^{-2}$\AA$^{-1}$. In order to determine peak positions and half widths (FWHM), the
experimental spectra were numerically fitted. Details of the fitting procedure can be found in ref.~\cite{Langer10}.

\section{Results}
As we have already mentioned in recent work \cite{Tege11} that it is not possible to parametrize the observed losses by a single loss peak due to their asymmetric form, and it seems that also in such a purely two-dimensional
system more than one collective electronic excitation is possible.

This is particularly obvious
for graphene on Ir(111). In fig.\ref{FIG1} we show a  comparison of loss spectra for monolayer graphene grown on Ir(111) and on the Si face of SiC(0001), which  have been measured for scattering vectors ranging from zero up to 0.20 \AA$^{-1}$. All spectra were measured at room
temperature at a primary energy of 20 eV. On the Ir(111) surface we found apart from asymmetric peak shapes at low loss energies and small $k_\parallel$ values two separable loss peaks that can clearly be discriminated at $k_\parallel \ge 0.07$ \AA$^{-1}$, as seen in fig.~\ref{FIG1}a).
The more intense peak is assigned to the monopole sheet plasmon. Compared to the plasmon
of graphene on SiC(0001) (see fig.~\ref{FIG1}b)), the
graphene sheet plasmon on Ir(111) is by a factor of five less intense and the full width
at half maximum (FWHM) is twice as large (for a detailed analysis of the FWHM, see below).

\begin{figure*}[tb]
 \centering
 \includegraphics[width=0.8\textwidth]{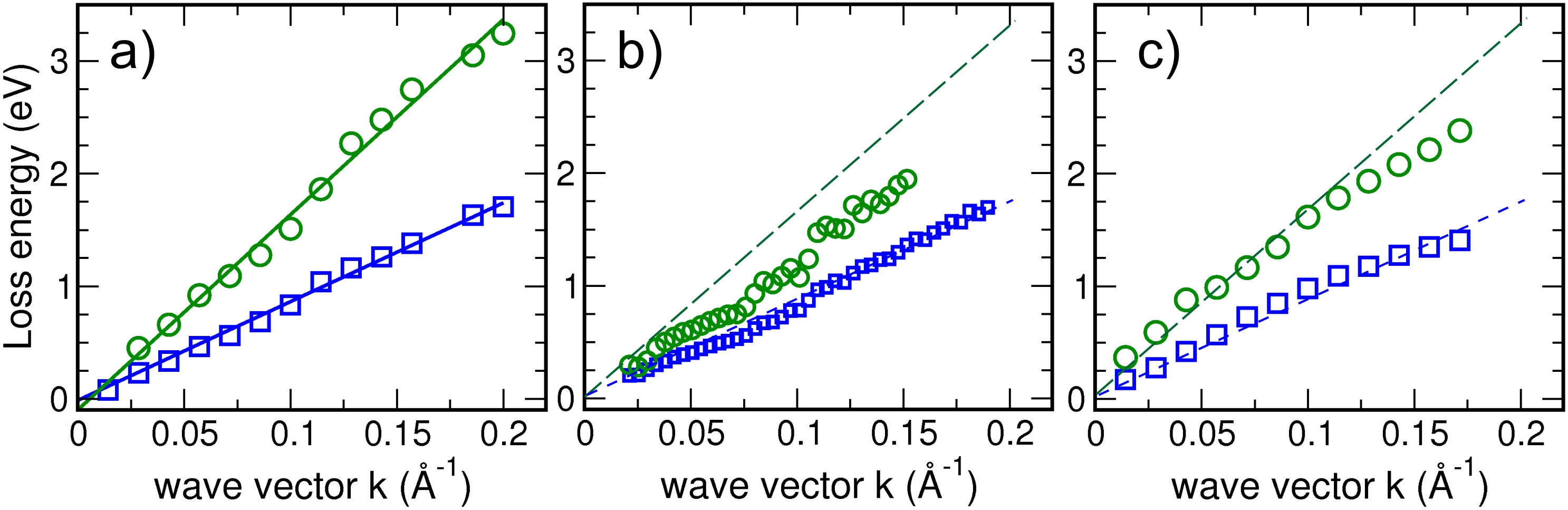}
 \caption{\label{FIG2}a) Plasmon dispersion of the two plasmonic branches of graphene on Ir(111).
b) same type of data evaluation for graphene on SiC(0001) (n-doped layer $n \approx 10^{13}$ cm$^{-2}$
c) Results for Na intercalated graphene on Ir(111). For comparison, we drew the average linear slopes of a) as
dashed lines in the other graphs.}
\end{figure*}

Nevertheless, the intensities of the second loss peak are higher than on SiC(0001). While the ratio
between the sheet plasmon and this high energy mode is 3:1 on SiC at $k_\parallel= 0.04$\AA$^{-1}$, it is close to 3:2 on Ir(111). Qualitatively, and partly even quantitatively, the same results were
obtained after adsorption of half a monolayer of Na. However, the intensity of the sheet plasmon increases by roughly a factor of five relative to the elastic peak, and the FWHM is reduced.
The intensity ratio between the two losses again increases to 3:1. This behavior is
expected for multipole plasmonic excitations when the charge distribution normal to the surface varies
due to different environments \cite{Liebsch97}.
With respect to the systems investigated here, the interaction of graphene with Ir(111)
is signicantly weaker than for SiC(0001), as directly obvious from ARPES measurements
\cite{Ohta06,Michely09}.
The Na layer turns out to be intercalated \cite{Langer11}. Therefore, it
is expected not only to change the electronic carrier concentration of graphene, but also
to further reduce the interaction between graphene and Ir(111).
Therefore, we assign the higher loss peak to a multiplasmon mode.
In fact, this multipole mode in a 2D sheet may be considered as the limiting case of
surface plasmons located at two interfaces and interacting with each other \cite{Liebsch97,Quinn92}.

These findings motivated us to further analyze our ELS-LEED data of graphene on SiC(0001) and to fit the data
with the same routine and the same procedures as for the data on Ir(111), and to describe the asymmetric shape observed there by a second Gaussian peak plus a residual background.
The results for some of the EELS spectra of this system are plotted in
fig.~\ref{FIG1}b). It is quite obvious also in this case that a single loss peak is insufficient to describe the shape of the observed loss peaks reasonably well. In addition, there is a systematic increase of separation between the two fitted peaks as a function of $k_\parallel$ in both systems.

Even more informative are plots of the peak positions of the losses as a function of $k_\parallel$, as shown in fig.~\ref{FIG2}. In fig.~\ref{FIG2}a), the plot of plasmon dispersion of graphene on Ir(111), we see that the
dispersion of both branches is strictly linear within error bars.  Therefore, also for the multipole mode
the excitation energy seems to vanish in the limit of long wavelengths:
$\omega \rightarrow 0$ for $k_\parallel \rightarrow 0$. This identifies also this second mode as a truely
two-dimensional mode with a vanishing restoring force in the limit of $k_\parallel \rightarrow 0$.

The ratio between the slopes, i.e. between
group velocities of both branches is $2\pm 0.05$. When we now turn to the graphene system on SiC(0001),
again both branches are detected (see fig.~\ref{FIG2}b),
but now the energy ratio between these branches is $1.4 \pm 0.1$.
Interestingly, the small dip at $k_\parallel = 0.08$\AA$^{-1}$ exists in both branches. A similar evaluation
on the Na intercalated Ir(111) surface yields an energy ratio of $1.6 \pm 0.07$. Qualitatively, and
even semi-quantitatively, the dispersion of the multipole branch follows closely that of the monopole mode.

Therefore, we will shortly discuss the monopole mode here. A more extended description and discussion can be found
in refs.~\cite{Tege11,Langer11}. The most remarkable result concerning the monopole plasmon dispersion is its
constancy of slope, when substrate and carrier concentration in the conduction band are changed. The average slope changes by 15\% at most, although the doping concentration is altered by at least three orders of magnitude from essentially non-doped in graphene on Ir(111) to strongly doped in the Na intercalated system  \cite{Langer11}. The doping level on SiC(0001) is in between these two extremes \cite{Tege11}.  This result clearly contradicts
the standard theories of a 2D electron gas \cite{Stern67, Hwang07}, which predict a significant dependence
on carrier concentration, but also on effective masses and on the environment. It would be highly accidental,
therefore, if these parameters would compensate each other so perfectly, in particular for graphene on Ir(111) with
and without Na intercalation. It seems that in this gap-less electronic system an effective integration over
occupied electronic states over a range of roughly 1. 5 eV takes place yielding an essentially constant
{\em effective} density of electrons participating in plasmonic excitations \cite{Langer11}, possibly
caused by the short plasmonic life times in these systems. However, while experimental parameters like step densities on the substrate etc. may modify details of the dispersion curves  or half widths of the observed losses, they
only act as second order corrections with this effective density still essentially fixed.

One of these parameters is the exact linearization of the dispersion curves observed on Ir, but not on SiC,
which can be understood by the screening of the plasmonic excitation by the bulk metal \cite{Pitarke04}, and has
been observed for 2D plasmons induced by partially filled surface states on Be(0001) \cite{Diaconescu07} and
Cu(111) \cite{Pohl10}.

\begin{figure*}[tb]
 \centering
 \includegraphics[width=0.8\textwidth]{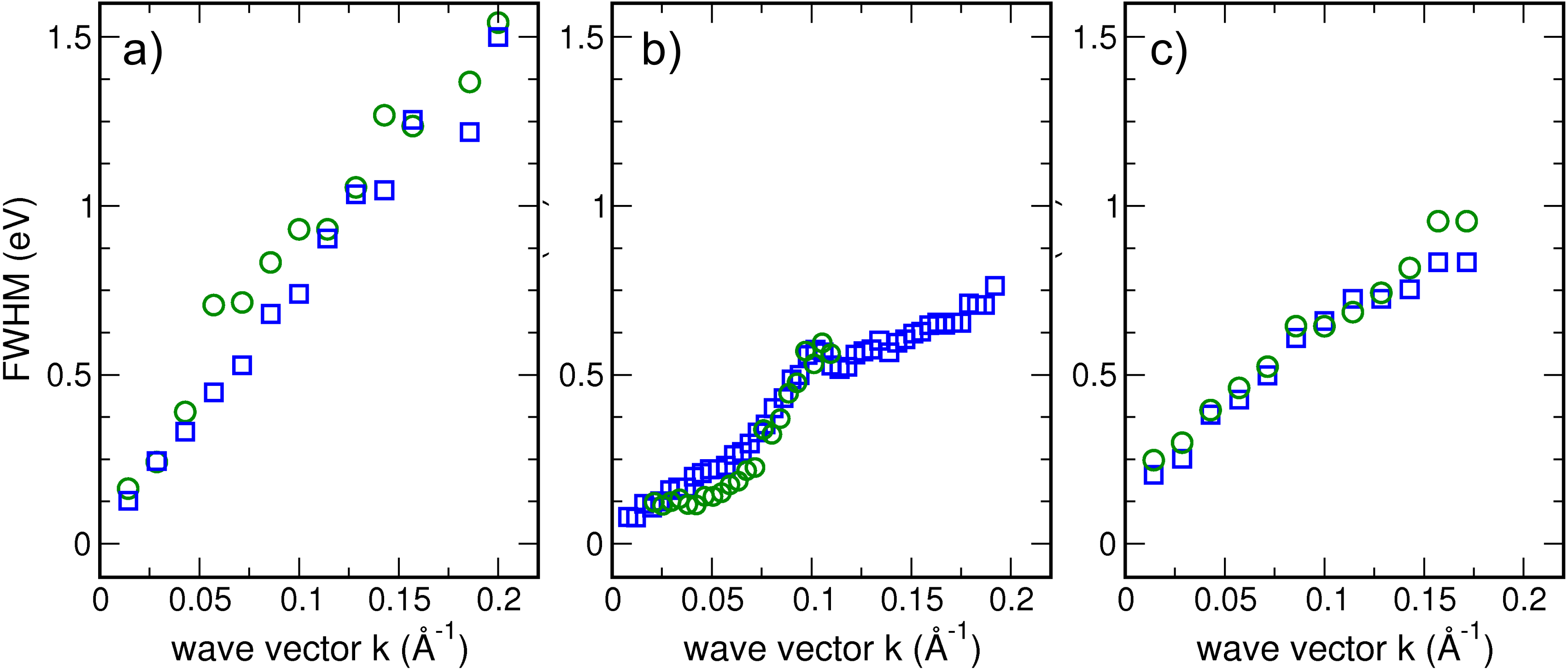}
 \caption{\label{FIG3} Half widths (FWHM) of loss peaks as a function of $k_\parallel$, a) for graphene on Ir(111),
b) for graphene on SiC(0001), c) for Na intercalated graphene on Ir(111). Squares denote data for the monopole
mode, circles those for the multipole mode.} .
\end{figure*}

As mentioned, the dispersion of the multipole mode shows essentially the same k-dependence as the monopole mode.
This close relationship may be understood considering the restricted possibilities of collective displacements
within a 2D electron gas parallel and normal to the plasmon sheet. Especially electronic collective
displacements normal to the graphene sheet, however, should be susceptible to the coupling between
graphene layer and the substrate in addition to the shielding by the substrate, since coupling to the substrate
modifies the restoring forces. In agreement with this expectation, we find that the energetic ratio between
monopole and multipole mode varies with substrate and in particular with the strength of interaction.
Considering these ratios for the three systems investigated here, however, they obviously
depend further on the details of interaction and spatial
electron distribution, not just on the strength of interaction. On the other hand, the sensitivity of the
energetic ratio to these parameters is a further indication for the existence of the multipole plasmon.

Coming back to the quantitative properties of the multipole plasmon dispersion, we observed even quantitative simultaneous changes of both dispersion curves as a function of $k_\parallel$,
e.g. the appearance of the dip in both plasmon dispersion curves on SiC(0001), can be
understood as being of the same origin. In this case, the dip was interpreted as a  coupling resonance between plasmonic and single particle excitations due to interband transitions \cite{Tege11}. Considering the band structure of graphene, the crossing of the multipole plasmon dispersion curve and the interband transition regime is only marginally shifted to smaller $k_\parallel$ values compared to the monopole mode, in agreement with the results shown in fig.~\ref{FIG2}b) and with the corresponding resonant increase of the FWHM [see fig.~\ref{FIG3}b)].

Generally, we find that the FWHM of monopole and multipole plasmons not only exhibit the same functional dependence as a function of $k_\parallel$, they even agree within error bar with each other. This is remarkable since with their significantly higher excitation energy, the phase space for electron
hole decay is much larger for the multipole plasmons and shorter life times than for monopole plasmons are expected.
This is obviously not the case and indicates that the same decay mechanisms must be effective for both modes, which, however, must be of other origin.

An alternative comes from our analysis of the monopole plasmon loss and its coupling with structure on Ir(111)
\cite{Langer11}. Here we found that the life time of the monopole plasmon on the Ir surface
is fully determined by the modulation of the graphene layer due to its interaction
with the substrate. In other words, although graphene is less strongly coupled to the Ir surface
than to SiC, so that graphene overgrows the substrate stepRocca07s in a carpet-like mode, this modulation is still sufficient to act as efficient source of momentum transfer. Only momentum transfer, however, is necessary for the transformation of the plasmonic excitation to an electron-hole excitation within graphene. Both the modulation by the Moir\'e and the step density contribute here \cite{Langer11}. From these findings we conclude that this mechanism with essentially the same probability is also effective for the multipole plasmons.

A similar situation was found  for the Na intercalated system and long plasmonic wavelengths.
Only in this systems some deviations between the FWHMs of monopole and multipole modes at the shortest
wavelengths ($k_\parallel > 0.12$\AA$^{-1}$) were found. For the monopole mode, we argued in  \cite{Langer11}
that the reduced moir\'e-like modulation of the graphene film in presence of Na intercalation increases
the life plasmonic life times compared to the situation without Na. This effect is obviously less
pronounced for the multipole mode, although the life times are still significantly longer than for graphene
on the pure Ir(111) surface.

\section{Conclusions}
We have demonstrated that  in a strictly two-dimensional electron system several plasmon
modes can be excited, and  have identified by energy and momentum resolved electron
loss spectroscopy two of these modes for
adsorbed single graphene sheets. These were called monopole and multipole plasmonic excitations.
The ratio of excitation energies at constant wavelength seems to depend on the type of
substrate (metal or semiconductor/insulator) and on the coupling strength. More effective
screening of the excitation by a metallic
substrate seems to increase the frequency of the multipole plasmon. Reduced coupling between substrate and graphene, on the other hand, decreases this ratio. Since reduced binding to a substrate is usually coupled with a change
in average bond lengths, evidenced here by an effectively reduced modulation of the
graphene layer in the Na intercalated Ir system, the electronic spillout at the interface
into vacuum is modified. Multipole modes are sensitive to the form of the decay of the electron density normal to the surface. This explains the
observed change of loss ratios when the coupling to the substrate is modified.

Regarding the ratios of intensities, there should be clearly a dependence on the oscillator strength of the
various modes and on their coupling to the multipole field of the incident electrons in an EELS experiment.
In addition, however, there is also a strong modulation of intensities by the dynamic electron backscattering
probability, since this probability varies strongly with incident or backscattered electron energy
and with their angle of incidence, particularly at low energies below 20 eV.
As an extreme, the observability of certain modes may be suppressed at given primary electron energies.

The overall insensitivity of the measured dispersion curves to the actual doping level of graphene is not
understood at present. It clearly invalidates the simple picture that only the electron density of the conduction bands participates in the plasmonic excitation. Although doping induces modifications of the
band structure close to the Dirac point, which may in part be responsible for compensational effects, our experiments show that the {\em effective} electron density 
is essentially independent of doping concentration.

The extreme confinement of the electronic excitations along a two-dimensional sheet turns out to make
these excitations extremely sensitive to all kinds of distortions of their 2D periodicity. Among these
distortions are variations of coupling to the substrate by steps and the associated 3D modulation of the
graphene film. Our results suggest that even for the most perfect substrate of SiC with
average terrace lengths of more than 100 nm there is still a dominance of substrate
imperfections for plasmon decay. Therefore, there is an urgent need for even more perfect substrates in order
to be able to study 2D plasmon physics in more detail.

\section{Acknowledgements}
We thank the group of T. Michely for lending an Ir(111) crystal to us, and for their help in
preparation of graphene layers.

\bibliography{plasmon-graphene}

\end{document}